# The origin of Jupiter's Great Red Spot

Agustín Sánchez-Lavega[1], Enrique García-Melendo[2], Jon Legarreta[1], Arnau Miró[3], Manel Soria[2], Kevin Ahrens-Velásquez[2]

[1] Escuela de Ingeniería de Bilbao, Universidad del País Vasco UPV/EHU, Bilbao, Spain

[2] Escola Superior d'Enginyeries Industrial, Aeroespacial i Audiovisual de Terrassa, Universitat Politècnica de Catalunya, Terrassa (Barcelona), Spain

[3] Barcelona Supercomputing Center, Barcelona, Spain

Corresponding author: first and last name (agustin.sanchez@ehu.eus)

**Key Points:**

- A study of historical observations suggests that Jupiter´s Great Red Spot was not the Permanent Spot reported by G. D. Cassini in 1665

- The temporal evolution of the shrinkage rate, area and eccentricity of the Great Red Spot and its Hollow have been precisely determined

- Observations and numerical simulations indicate that the genesis of the Great Red Spot was due to a disturbance in Jupiter's sheared flow

**Abstract**

Jupiter's Great Red Spot (GRS) is the largest and longest-lived known vortex of all solar system planets but its lifetime is debated and its formation mechanism remains hidden. G. D. Cassini discovered in 1665 the presence of a dark oval at the GRS latitude, known as the "Permanent Spot" (PS) that was observed until 1713. We show from historical observations of its size evolution and motions that PS is unlikely to correspond to the current GRS, that was first observed in 1831. Numerical simulations rule out that the GRS formed by the merging of vortices or by a superstorm, but most likely formed from a flow disturbance between the two opposed Jovian zonal jets north and south of it. If so, the early GRS should have had a low tangential velocity so that its rotation velocity has increased over time as it has shrunk.

**Plain Language Summary**

Jupiter's Great Red Spot (GRS) is probably the best known atmospheric feature and a popular icon among solar system objects. Its large oval shape, contrasted red color and longevity, have made it an easily visible target for small telescopes. From historical measurements of size and motions, we show that most likely the current GRS was first reported in 1831 and is not the Permanent Spot observed by G. D. Cassini and others between 1665 and 1713. Numerical models show that the GRS genesis could have taken place from an elongated and shallow, low speed circulation cell, produced in the meridionally sheared flow.

**1 Introduction**

The GRS is a giant anticyclone vortex that comprises two main regions as observed at optical wavelengths, a red oval (the GRS properly said), and an outer "whitish area" surrounding it, more extended along its northern part, known as the Hollow (Peek, 1958; Rogers, 1995; Ingersoll et al., 2004) (Figure 1b). Its visibility changes depending on contrast with surrounding clouds and sometimes it manifests as a single clear oval, covering both areas (red oval and Hollow). Wind measurements from cloud motions show that the Hollow edge outlines the boundary of the circulation associated with the vortex, and thus the red oval and its Hollow fully comprises its dynamical area (Mitchell et al., 1981; Choi et al., 2007; Asay-Davis et al., 2009; Sánchez-Lavega et al., 2019; Wong et al., 2021).

The formation mechanism that gave rise to the GRS is unknown. And its longevity is a matter of debate, and to date it is not clear if the GRS was the dark oval, nicknamed Permanent Spot (PS), reported by Giovanni Domenico Cassini and others from 1665 to 1713 (Cassini, 1666; Chapman, 1968; Falorni, 1987; Rogers, 1995; Hockey, 1999; Simon, 2016; Chapman, 2016) (Fig. 1, Fig. S1). In order to clarify the relationship between PS and the GRS, we first present an in deep analysis of all the available observations of PS and the GRS, particularly up to the 20$^{th}$ century. Then, we study and compare a year-by-year measurement of their size, ellipticity, area and motions, as well as of the Hollow area, from the earliest available observations and until 2023. This study extends and complete the results previously presented by Beebe and Youngblood (1979), Rogers (1995) and Simon et al. (2018), and makes it possible to specify the relationship between PS and the GRS-Hollow.

In a second part of this work, guided by these historical observations and the recent data on the GRS, we present numerical simulations of different dynamical mechanisms that could have lead to the genesis of the GRS. We explore three plausible scenarios: a "super-storm", the mergers of vortex chains smaller than the GRS, and its birth as an elongated cell (a proto-GRS) generated by a disturbance in the meridionally sheared zonal winds.

## 2 Methodology

### 2.1 Image measurement

The appearance of the GRS and its Hollow throughout the history of Jupiter observations has been highly variable due to changes in size, albedo and contrast with surrounding clouds (Peek, 1958; Rogers, 1995) (Fig. 1, Fig. S1-S3) We have updated and extended previous measurements of the GRS size and motion using the data sets from a large number of sources (a list is given in Table S1) and standardising the measurement methodology. Measurements of the size of the Permanent Spot (PS) have been performed on all the available drawings in the period 1665-1713, and of the GRS and clear oval and Hollow (these last two for the first time) on early drawings (1831-1879), photographs (1879-1980) and more recently on digital images (1980-2023). We have used the WinJupos (2024) software to navigate the images (i. e. fix the limb, the terminator, and the coordinates on the planet). Whenever possible, we have used photographs taken in blue-violet filters where the limb contrast improves, reducing errors. Measurements of the East-West size have uncertainties of ±5° in the drawings from 1831-1878 and in the range ±0.5° to ±2° in the photographic records depending on the image resolution (1° in longitude = 1151 km). For ground-based digital images the precision is ±0.5° and for HST is ±0.1°. We have relied on previously published data for observations made from flyby and orbiting space missions in Jupiter (Simon et al., 2018).

The yearly zonal velocity of the center of the GRS was derived from the published rotation periods for the period 1890-1948 (Peek, 1958) and from data taken from the different sources in Table S1. In addition, the velocity was calculated directly from the measurments of its yearly longitude position and retrieved drift rate between consecutive years (difference in longitude divided by the time interval). Since these velocities represent an average over a period ~ one year, the velocity error is small, in the range ± 1 ms$^{-1}$.

### 2.2 Numerical simulations

We have used two dynamical models to perform simulations of the GRS genesis. Since the estimated length is likely to be much greater than the depth of the current GRS as determined from Juno spacecraft studies (Bolton et al., 2021; Parisi et al., 2021) and theoretical models (Vasavada and Showman, 2005; Read, 2024), we use a Shallow Water (SW) model (García-Melendo and Sánchez-Lavega, 2017; Soria et al., 2023) and the Explicit Planetary Isentropic Coordinate (EPIC) (Dowling et al., 1998) operating for Jupiter conditions. The SW model calculates the evolution of the potential vorticity PV (s$^{-1}$) in a meridionally sheared flow in a mono-layer with a domain extending in longitude from 0° to 180° and from 5°S to 45°S in latitude. The spatial resolution ranges from 0.2°/pixel to 0.02°/pixel per control volume with time step Δt = 0.25-10 s. The background sheared wind profile comes from Hueso et al. (2017). We use for post-processing visualization the software Paraview (2024). To consider the possible effects of vertical stratification in the temperature and wind velocity of the background atmosphere, we performed additional simulations with the EPIC code that solves the hydrostatic primitive equations using potential temperature as the vertical coordinate and computes the evolution of potential vorticity (units PVU $10^{-6}$ K kg$^{-1}$ m$^2$ s$^{-1}$) (Dowling et al., 1998). This model has been amply tested and extensively used in the study of Jupiter disturbances and the GRS (García-Melendo et al., 2005; Legarreta and Sánchez-Lavega, 2008; Sánchez-Lavega et al., 2008; Morales-Juberías and Dowling, 2013; Iñurrigarro et al., 2020). The evolution of the flow field is mapped following the introduction of a localized disturbance (a mass injection source in SW, heat injection in EPIC and anticyclones in both models). Details of the simulation parameters we have used are given in the corresponding sections, the figure captions and compiled in Tables S2 and S3.

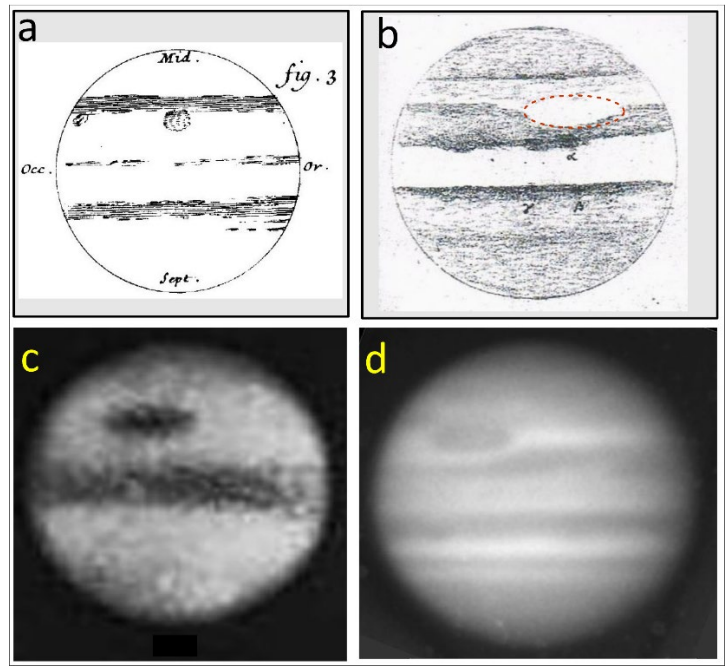

**Figure 1.** *The Permanent Spot and the early Great Red Spot. (a) Drawing of the Permanent Spot (PS) by G. D. Cassini, 19 January 1672. (b) Drawing by S. Swabe in 10 May 1851, showing the GRS area as a clear oval with limits marked by its Hollow (draw by a red dashed line). (c) Photograph by A. A. Common obtained in Ealing (London) on 3 September 1879 using a 91 cm reflector (5.30 m focal length, 1 sec exposures) (Clerke, 1887). The GRS shows prominently as a "dark" oval due to its red color and photographic plate sensibility to violet-blue wavelengths. (d) Photograph from Observatory Lick with a yellow filter on 14 October 1890. All figures show the astronomical view of Jupiter (South up, East left) to preserve notes on the drawings.*

**3 Analysis of PS, GRS and Hollow data**

    3.1 The Permanent Spot and the GRS early observations

The Permanent Spot was first reported by G. D. Cassini and other astronomers in July-September 1665 (Cassini, 1666; Falorni, 1987). It has been shown that a spot previously reported by Robert Hooke in 1664 was not the PS (Falorni, 1987; Rogers, 1995; Hockey, 1999). However, PS could have been observed even earlier by L. Bandtius, on 2 November 1632, who reported the presence of an oval approximately one-seventh the size of Jupiter's radius (Riccioli, 1665; Graney, 2010). PS was subsequently observed by Cassini and others in 1667, 1672, 1677, 1685-87, 1690-91, 1694, 1708, and was last reported in 1713 by M. Maraldi (Cassini, 1672; Cassini, 1692; Maraldi, 1708; Rogers, 1995). This indicates that the lifetime of PS was at least ~ 81 years. In none of these observations is any color of PS mentioned. However, a painting of Jupiter in 1711 intriguingly shows PS with a red tint (Hockey, 1999; Johns, 1992), remembering the current GRS (Fig. S2).

No reports of PS or any sign of its presence exists in the available observations of Jupiter between 1713 and 1831, a long period of ~ 118 years (Rogers, 1995; Hockey, 1999). Examination of the drawings by renowned astronomers of the epoch as M. Messier in 1769, W. Herschel in 1778, H. Schroeder in 1785-86 and others, shows belts and isolated spots, but in no case a PS or a similar spot at its latitude confirming previous findings (Messier, 1769; Herschel, 1781; Rogers, 1995; Dobbins et al., 1997; Hockey, 1999). It would be surprising if, had it existed, none of the astronomers of the time had reported PS. Considering the small size of PS in the drawings in 1672-1692, it is most likely that this lack of observations during such a large period means that PS disappeared. The first drawings showing the signature of the current GRS, recorded by its Hollow, date back to 1831, and drawings in the 1870-71 showed it as a well-defined clear oval enclosed by a dark elliptical ring (Rogers, 1995; Hockey, 1999) (Fig. 1b, Fig. S3). This oval became reddish and surrounded by the Hollow in ~ 1872-1876 (Rogers, 1995, Fig. S2). The first available photograph showing a prominent GRS was obtained in 1879 (Clerke, 1887, Fig. 1c). The current GRS has therefore certainly been in existence for 193 years.

3.2 Sizes and motions of PS, GRS and its Hollow

We have measured the size of PS, the red oval (GRS) and the Hollow (and "clear oval" as it shows in some cases) from 1665 to present. Fig. 2 shows their length in the zonal direction (east-west) and their width in the meridional direction (north-south) (Sánchez-Lavega, 2024). The length of PS is 2-3 times smaller than that of the 1879 GRS. The length of the GRS decreased over time at an average rate of -0.18°/year (207 km/year) (increasing in last years to -0.3°/year), in agreement with previous results for shorter time periods of analysis (Beebe and Youngblood, 1979; Simon et al., 2018). The GRS experienced a transient increase in length from ~ 1927 to 1939 at a rate of +0.07°/year (80 km/year), when it engulfed clouds from a large and enduring South Tropical Disturbance (STrD) that developed at the time (Rogers, 1995). The Hollow followed a similar average shrinkage rate of -0.20°/year (230 km/year). Within the inaccuracy inherent to measure the drawings, also PS seems to show a similar decrease in length. The extrapolation back in time of a polynomial fit to the shrinkages of both the GRS and Hollow strongly suggests that PS is not the GRS (Fig. 2a). PS would have had to grow steadily from 1713 to 1879 at a rate of ~ +0.14°/year (160 km/year) to be the GRS. This is highly unlikely since, as shown above, no reports of PS or GRS exist during this long period and, in addition, no continuously sustained grow in size has been never reported in Jupiter's vortices (Rogers, 1995; Ingersoll et al., 2004).

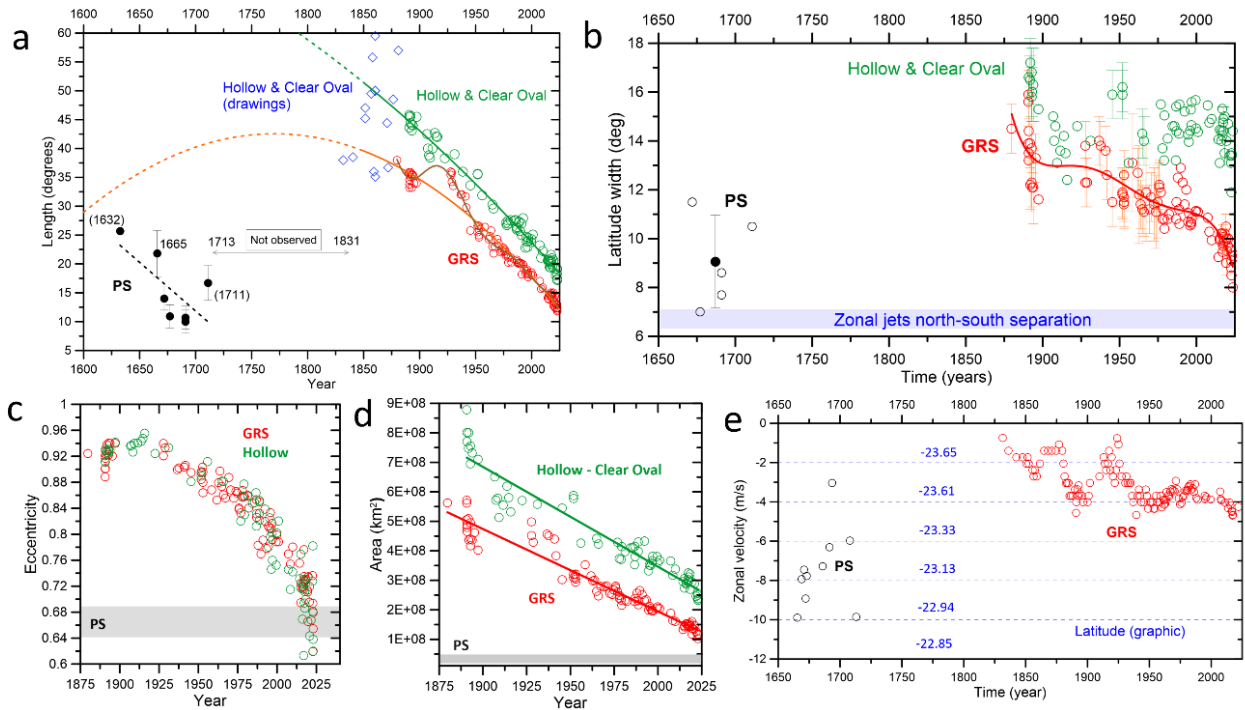

**Figure 2.** *Temporal evolution of the sizes and motions of PS, the GRS and its Hollow. (a) Measured zonal length (longitude width) of PS (black circles, 1665-1711, parenthesis indicate a large uncertainty), GRS (red circles, 1879-2023) and Hollow and Clear Oval (green circles from photographs, 1890-2023, and blue diamonds from drawings, 1831-1891). The period without PS or GRS reports is indicated (1713-1831). The dashed black line shows a linear fit to a possible decreasing trend in length for PS. The red and green continuous lines show a two-degree polynomial fit to the decrease in zonal length of the GRS and Hollow – Clear oval morphology. The brown line is a ten-order fit to try to capture the transient increase in length around ~1927-1939. The red and green dashed lines are the extrapolations back in time of the fits which shows that it is highly unlikely that the current GRS to be the former Cassini's PS. (b) Meridional width evolution with symbols and dates as in* (a). *The black dot show the mean width for Cassini's PS. The red line shows a five-degree polynomial fit to the decrease in zonal width of the GRS. The blue band shows the meridional distance between the velocity peaks (eastward-westward) of the zonal jets north and south the GRS center. (c) The decrease of the GRS and Hollow eccentricity. (d) The decrease of the GRS and Hollow areas and linear fits. The horizontal grey band shows the range for PS eccentricity and area as measured from drawings between 1665 and 1713 (Fig. S1). (e) Zonal velocity of PS and GRS. The dashed horizontal lines mark the planetographic latitudes where the zonal velocities match. The data are available in Sánchez-Lavega (2024).*

In the meridional direction, the GRS gradually decreased its width since 1879 at a mean rate of -0.03°/year (36 km/year). The Hollow width exhibited a fluctuating but a global decrease at a mean rate -0.09°/year (11 km/year). Note that the shrinkage of both has accelerated since 2010 to -0.17°/year and currently, the GRS has about the same width than PS (Fig. 2b), close to the distance separating the peak of the zonal jets north and south of the GRS (Simon et al., 2018;

Sánchez-Lavega et al. 2021). Assuming that the GRS and Hollow are ellipses with semi-axes ($a$, $b$), their eccentricity $e = \sqrt{1-(b/a)^2}$ decreased from ~ 0.92 in 1879 to 0.6 in 2023, i.e. both are becoming rounded-shape ovals (Fig. 2c). Their area $A = \pi ab$ decreased approximately linearly (Fig. 2d) and if this shrinkage persists, it could either lead to one of two cases: the GRS disappearance (as was the case of PS) or the GRS reaching a stable long-lived size. Note also that the eccentricity and area of the current GRS are similar to that of PS. As a reference, the eccentricity and area of the current red oval are similar to that of PS (Fig. 2c-d).

The zonal velocity drift of PS ranged from u ~ -10 ms$^{-1}$ to -6 ms$^{-1}$ and that of the GRS from ~ -4 ms$^{-1}$ to -1 ms$^{-1}$ (Fig. 2e) (Sánchez-Lavega, 2024). This velocity difference may be due to a shift in latitude of their centres by no more than 1° (relative to the background zonal wind profile), or be intrinsic and related to their dynamical properties, or to a combination of both. This different velocity has been another argument used to indicate that PS is not the GRS (Rogers, 1995; Simon, 2016).

## 4 Numerical simulations results

### 4.1 A Super-storm mechanism

On Saturn, convective storm outbreaks in anticyclone sheared flows generate anticyclone oval vortices (Dyudina et al., 2007). A significant case was the recent great storm (the Great White Spot GWS 2010) that generated an anticyclone that still lasts today (Sayanagi et al., 2013; Sánchez-Lavega et al., 2018). We study whether the GRS could have been generated in a similar way by an energetic moist convective "super-storm" on Jupiter. We have performed numerical simulations of the response of the Jovian flow at the GRS latitude (~22°S to 24°S) to a localized Gaussian heat injection in EPIC (García-Melendo et al., 2005; Iñurrigarro et al., 2022) and to a mass injection in the SW (García-Melendo and Sánchez-Lavega, 2017; García-Melendo et al., 2013). Our simulations generate a single oval anticyclone (Fig. 3a-b, Fig S4-S5) but its length is always smaller than the early GRS (Fig. 1c-d, Fig. 2). Increasing the intensity and the size and duration of the energy and mass injections produce unrealistic round oval shapes and rotation velocities much higher than those observed in the current GRS. It has also been proposed that anticyclones could be generated by deep convection driven by the internal energy of Jupiter but the published simulations do not resemble the early GRS (Cai et al., 2022). In any case, such a type of simulated super-storm has never been observed at the latitude of the GRS.

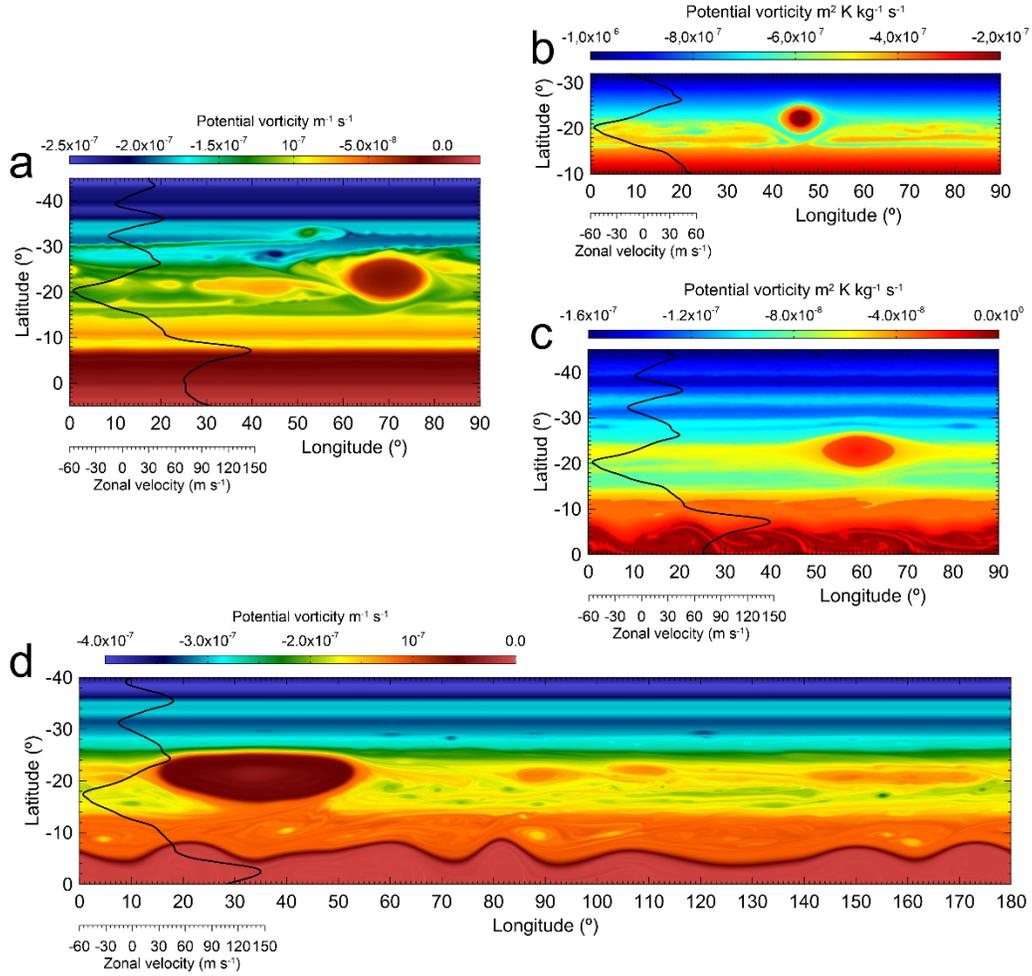

**Figure 3.** *Numerical simulations of the origin of the GRS from a Super-storm and Vortices mergers. Maps of potential vorticity PV in the SW and EPIC models (with units indicated for each case, Methods). (a) Superstorm in SW. A mass injection of 7 times $10^{11}$ $m^3 s^{-1}$ is introduced during 10 days in a Gaussian area with a radius of 7° at planetographic latitude 23°S, generating an anticyclone with a length to width 20° times 20° (Fig. S4). (b) Superstorm in EPIC. A heat impulse with a Gaussian shape with size 0.5° and intensity of 1.5 $Wkg^{-1}$ is introduced during 10 days at latitude 23.7°S, generating an anticyclone 9° times 6° with a Hollow-like with a size ~ 15° times 9° (Fig. S5). (c) Mergers of four anticyclones in EPIC with a size 8° times 7° and periphery velocity 120 $ms^{-1}$ located at latitudes from 22°S to 22.5°S (Fig. S6) resulting in an oval with size ~ 16° times 15°. The wind profile is to the left of each panel. (d) Mergers of five anticyclones in SW, four at 22°S and a fifth at 22.5°S with a size 15° times 10° and north and south periphery velocity V = 90 $ms^{-1}$ (Fig. S7) resulting in a single vortex with a size 41° times 12°.*

4.2 Anticyclone mergers

The merging of anticyclones is a well-known phenomenon in Jupiter (Ingersoll et al., 2004). Historically, the most relevant case was the merger of the three large and long-lived ovals

BC, DE, FA at 33°S that, after ~ 60 years of existence (Peek, 1958; Rogers, 1995), gave rise to the present-day single anticyclone oval BA (Sánchez-Lavega et al., 1999, 2001). We performed SW and EPIC numerical simulations of the merger of groups of up to 4-5 anticyclones in geostrophic balance centered at latitudes ranging from 19°S to 24°S. The merging anticyclones have initial sizes (East-West length x North-South width) from 8° times 7° to 15° times10° and peripheral velocities from $V$ = 90 to 120 ms$^{-1}$, typical of medium-large vortices in Jupiter. In all cases, the mergers form a new single and larger anticyclone than their precursors (Fig. 3c-d, Figs. S6-S7). However, both EPIC and SW simulations show that to form large anticyclones as the 1831-1879 early GRS, would require mergers of vortices as large as the current GRS, and in that case the resulting anticyclone has higher zonal rotational velocities than that currently observed in the GRS, something unexpected. Moreover, this kind of series of vortices or the disturbance producing them has never been observed at Jupiter, and if it had occurred, previous to the 1831 detection, it should have been reported due to its expected visibility.

### 4.3 The GRS genesis from a zonal flow disturbance

From 1831 to ~ 1877 the early GRS manifested as a Hollow and clear oval, with an east-west length ~ 50°-60° (Fig. 1b, Fig. S3). According to the measured shrinkage rate, it could have had a length ~ 70° in 1725 (Fig. 2a). This elongated cell could have been formed from a South Tropical Disturbance (STrD), an instability that initiates with the formation of dark curved meridional regions that act as barriers to the zonal flow (Rogers, 1995, 2008). North of the GRS at 20°S the velocity is u ~ -50 ms$^{-1}$ (westward, u <0) and South at 26°S is u ~ +40 ms$^{-1}$ (eastward, u > 0) (Fig. 4a-c). This flow becomes confined East-West by the two curved regions and North and South by the two jets. The initial peripheral velocity in the closed circulation cell would be that of the zonal jets, i. e. V ~ ±45 ms$^{-1}$. To test this hypothesis, we have performed simulations of the stability of long cells against different initial tangential velocities along their periphery. We introduced an elongated cell simulating the STrD as shown in Figures 4a-c, between the two opposing north and south jet streams. We have tested circulating cells with different lengths (between 45° and 80°), meridional widths (between 11° and 13°) and tangential peripheral along its border with velocities (50 – 120 ms$^{-1}$). Other ranges of the parameter values used in the simulations are specified in Tables S2 and S3. The sensitivity of the simulations to these parameters is obtained from a direct comparison between the PV maps (the size, stability and shape of the simulated vortex) with the observed GRS. The results show that these long cells are unstable if their initial velocity is that of the zonal jets, but gain in stability and robustness when the V > 50 to 75 ms$^{-1}$ (Fig 4d-e, Fig. S8). The East-West and North-South velocity profiles in the stable cell resemble closely those observed in the GRS (Choi et al., 2007; Sánchez-Lavega et al., 2021), with peak velocities that depend on the initial V introduced. We therefore propose that the GRS generated from a long cell resulting from the STrD, that acquired coherence and compactness as it shrank, increasing its peripheral zonal velocity to V ~ 70-100 ms$^{-1}$.

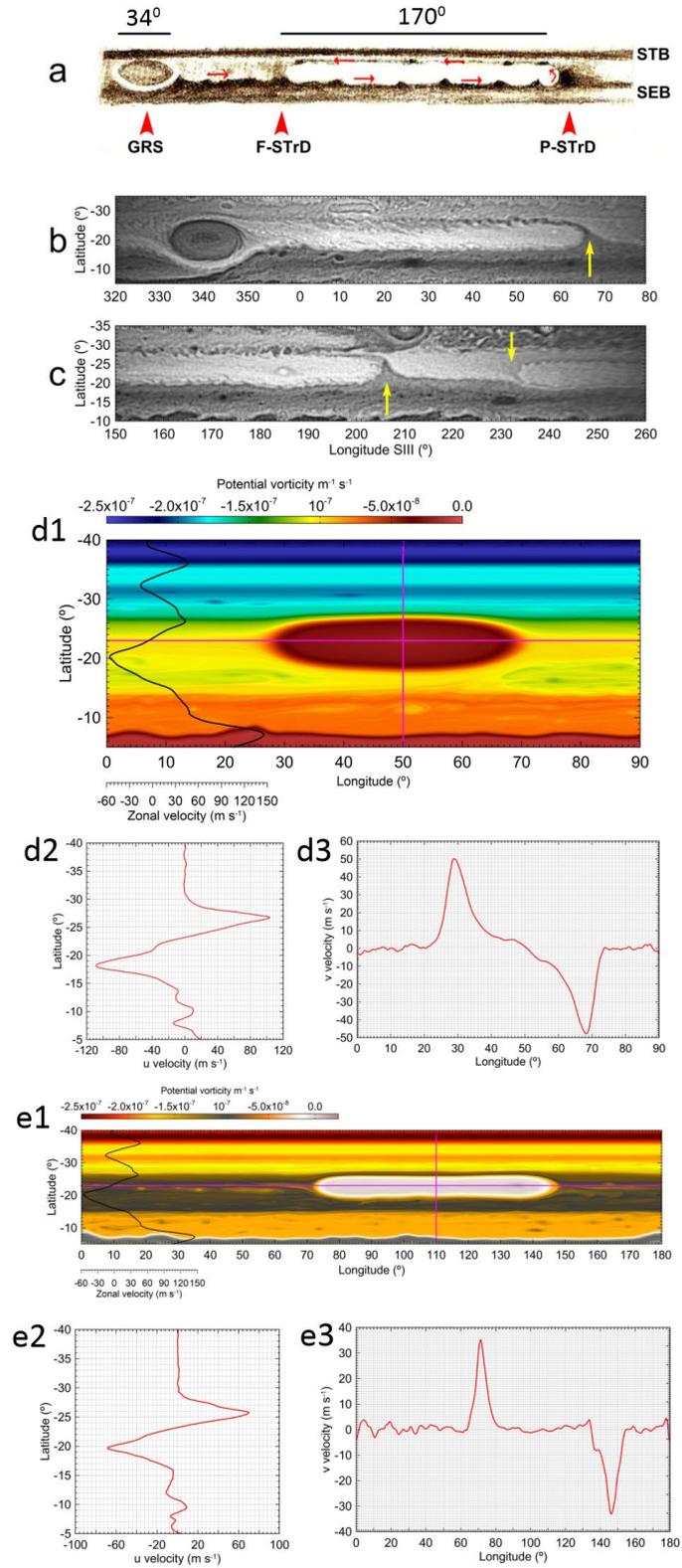
**Figure 4.** *Jupiter's South Tropical Disturbance and the GRS. (a)-(c) Strip maps of Jupiter showing the STrD curved dark areas (barriers) identified by arrows (P for preceding, F for following). (a)*

*Drawing by T. E. R. Phillips in 1931-32 of the STrD. The red arrows indicate the flow direction with the longitude scale indicated; (b) (c) Maps from images taken by the New Horizons spacecraft obtained during its Jupiter flyby in February 2007. The yellow arrows mark the position of the STrD "columns"; (d1) PV map in a SW simulation of the stability of a long cell with velocity at periphery $V = 100$ ms$^{-1}$ after 425.5 days simulation. The cell stable final size has 43° times 9.5° resembling the 1890 GRS (Fig. 1d); (d2) (d3) East-West and North-South velocity profiles across the cell center in (d1) marked by the red lines; (e1) Stability of a long-cell with a size ~ 75° times 8° and $V = 75$ ms$^{-1}$ at latitude 22.5°S, representing a STrD as a precursor of the GRS after 300 days of simulation; (e2) (e3) East-West and North-South velocity profiles across the cell center in (e1) marked by the red lines in (e1). Simulation data for (d1) and (e1) available in García-Melendo et al (2024).*

## 5 Conclusions

From these simulations, we conclude that the super-storm and the mergers mechanisms, although they generate a single anticyclone, are unlikely to have formed the GRS. Both phenomena have never been observed at the GRS latitude and, if they had occurred, astronomers at that time would have reported it. The elongated, slowly rotating cell, is reminiscent of the early observations of the GRS in mid-19$^{th}$ century. The STrD mechanism, which is a common disturbance at this latitude of Jupiter, seems more plausible to have generated a proto-GRS. A similar mechanism may have been behind the formation of Jupiter's other large and long-lived anticyclones (BC, DE, FA) located between two jets further south at 33°S. Finally, the comparison of the rotation speed of the GRS-precursor predicted by these models, with the recent measurements of the GRS circulation made by space missions (Wong et al., 2021), indicates that the GRS has been increasing its rotation speed in time as it shrunk, acquiring coherence and compactness, and forming the current rounder vortex.

**Data Availability Statement**

For the image measurements we used the WinJupos (2024) software (http://jupos.org/gh/download.htm). The Shallow Water code used in the numerical simulations is described in García-Melendo and Sánche-Lavega (2017) and can be downloaded in Soria et al. (2023). The EPIC code used in the numerical simulations is described in Dowling et al. (1998) and can be downloaded from NASA PDS, The Planetary Atmospheres Node
https://pds-atmospheres.nmsu.edu/data_and_services/software/epic/epic.htm
We used the visualization software ParaView (Paraview, 2024) to generate the images from code output (https://www.paraview.org/download/)
The long-cell SW simulations presented in figure 4 are available in García-Melendo et al. (2024) (https://doi.org/10.5281/zenodo.11120114).

**Conflict of Interest Statement**

The authors have no conflicts of interest to declare

**Supporting Information**

Figures S1 to S8

Table S1 to S3

**Acknowledgments**

ASL and JL were supported by Grupos Gobierno Vasco IT1742-22 and by Grant PID2019-109467GB-I00 funded by MCIN/AEI/10.13039/501100011033/. EGM, AM, MS and KAV thankfully acknowledge the computer resources provided by Red Española de Supercomputación (RES) under the projects AECT-2020-1-0005 and RES-AECT-2021-2-0009. We also acknowledge the Barcelona Supercomputing Center for awarding us access to the MareNostrum IV machine based in Barcelona, Spain and the Centro de Astrofísica de La Palma (IAC) for awarding us access to the LaPalma machine based in La Palma, Spain. We acknowledge Josep María Gómez-Forrellad for assisting in retrieving data from historical sources.

*Geophysical Research Letters*

Supporting Information for

# The origin of Jupiter's Great Red Spot

Agustín Sánchez Lavega[1], Enrique García-Melendo[2], Jon Legarreta[1], Arnau Miró[3], Manel Soria[2], Kevin Ahrens-Velásquez[2]

[1] Escuela de Ingeniería de Bilbao, Universidad del País Vasco UPV/EHU, Bilbao, Spain

[2] Escola Superior d'Enginyeries Industrial, Aeroespacial i Audiovisual de Terrassa, Universitat Politècnica de Catalunya, Terrassa (Barcelona), Spain

[3] Barcelona Supercomputing Center, Barcelona, Spain

**Contents of this file**

Text
Figures S1 to S8
Tables S1 to S3

**Introduction**

This Supporting Information contains images of the Permanent Spot and the old Great Red Spot (Figures S1 to S3), numerical simulations of the GRS origin by different mechanisms (Figures S4 to S8), and Table S1 with the sources and links for the images used in this study. Tables S2 and S3 give the input parameters for the numerical simulations presented in Figures 3-4 and Figures S4-S8.

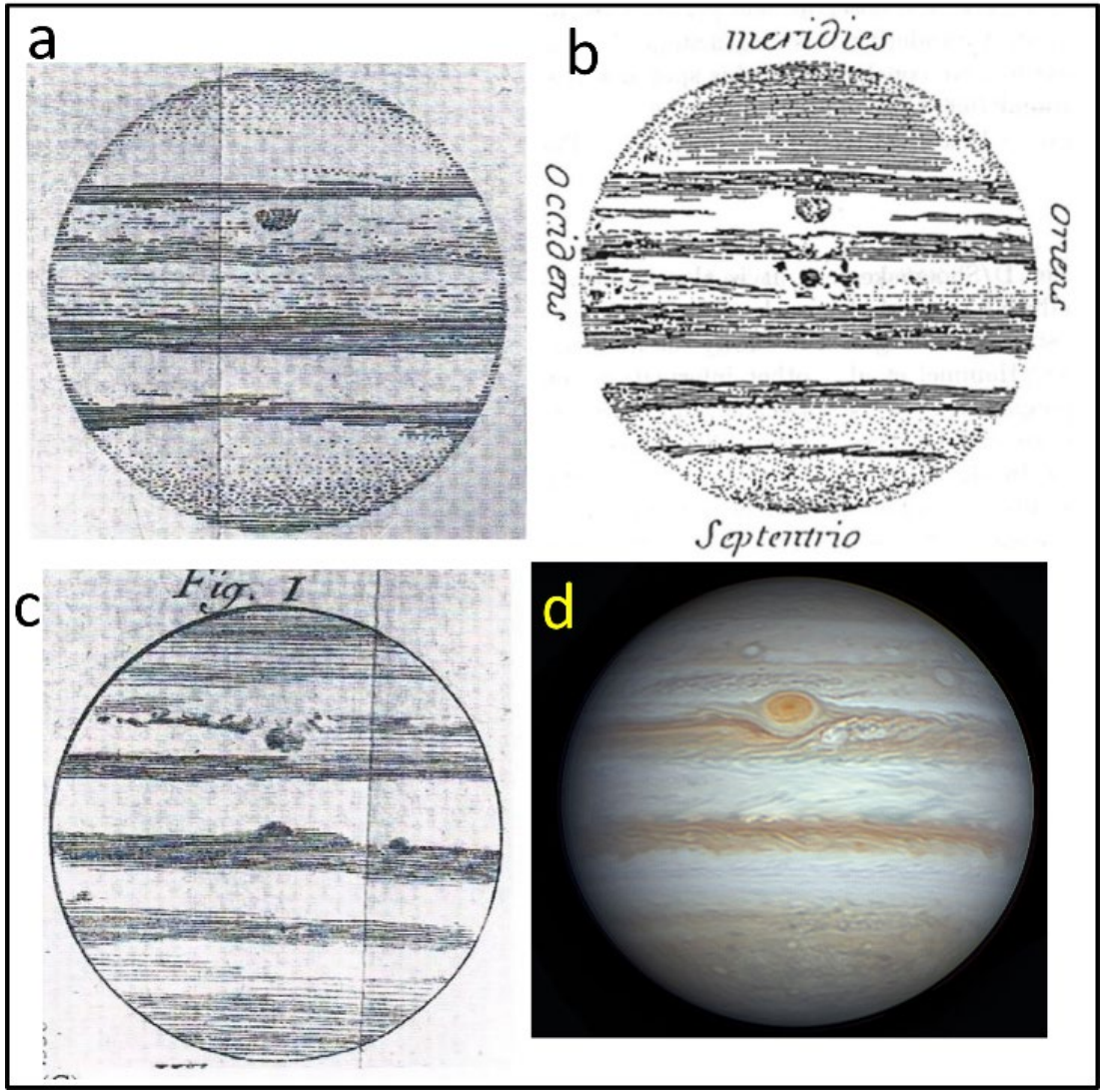

**Figure S1.** A comparison between the Permanent Spot (PS) and the current GRS. (a), (b), (c) show drawings by G. D. Cassini in July 1677 (a), December 1690 (b), January 1691 (c) (see also Fig. 1a for 19 January 1672). The PS size from these four Cassini drawings is 11.4° times 9.15°. Image (d) shows for comparison the current GRS in an image obtained on 10 August 2023 by Eric Sussenbach from Willemstad, Curacao (Dutch Caribbean). The GRS red oval has a length of 12.1° and width of 8° and the Hollow has a length of 18° and a width of 14.4°. The Jupiter disk has the same projected size in all these images.

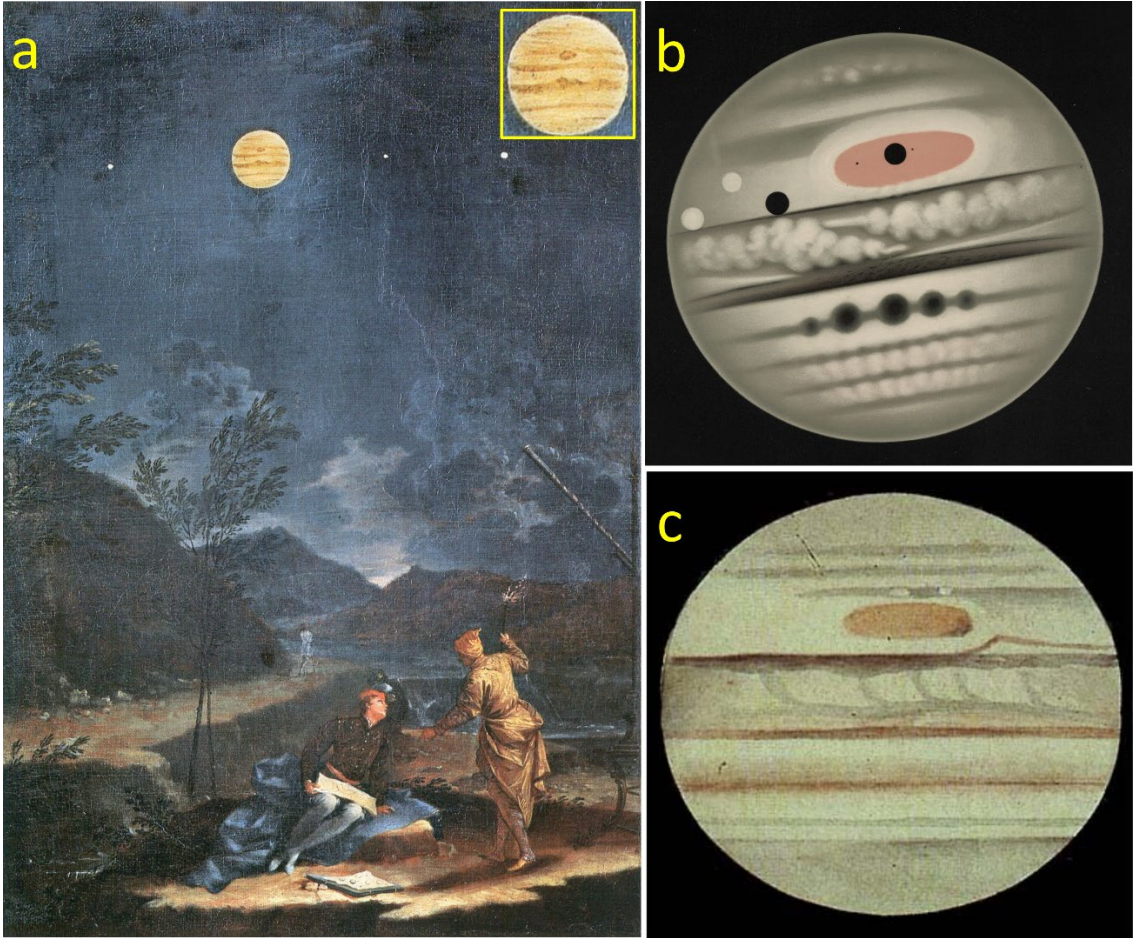

**Figure S2.** The red colour of Permanent Spot and the GRS described for the first time. (a) Paint of Jupiter by Donato Creti in 1711, showing a reddish PS (inset), as part of an astronomical paints series called "illustrated prospectus" (Roma, Pinacoteca Vaticana). Although the PS color may be a painter's license, it is more likely that G. D. Cassini or E. Manfredi, who inspired the painting, indicated such a color to D. Creti [Ref. 19]. (b) Drawing by French artist and painter E. L. Trouvelot on 2 November 1880 at ~ 02 h 20 min UT (with time corrected to the written date in the drawing using a time fit to the satellite position). (c) Drawing by T. G. Elger on 28 Novembre 1881. Note in (b) and (c) that the GRS is surrounded by an oval clear area corresponding to a well developed Hollow.

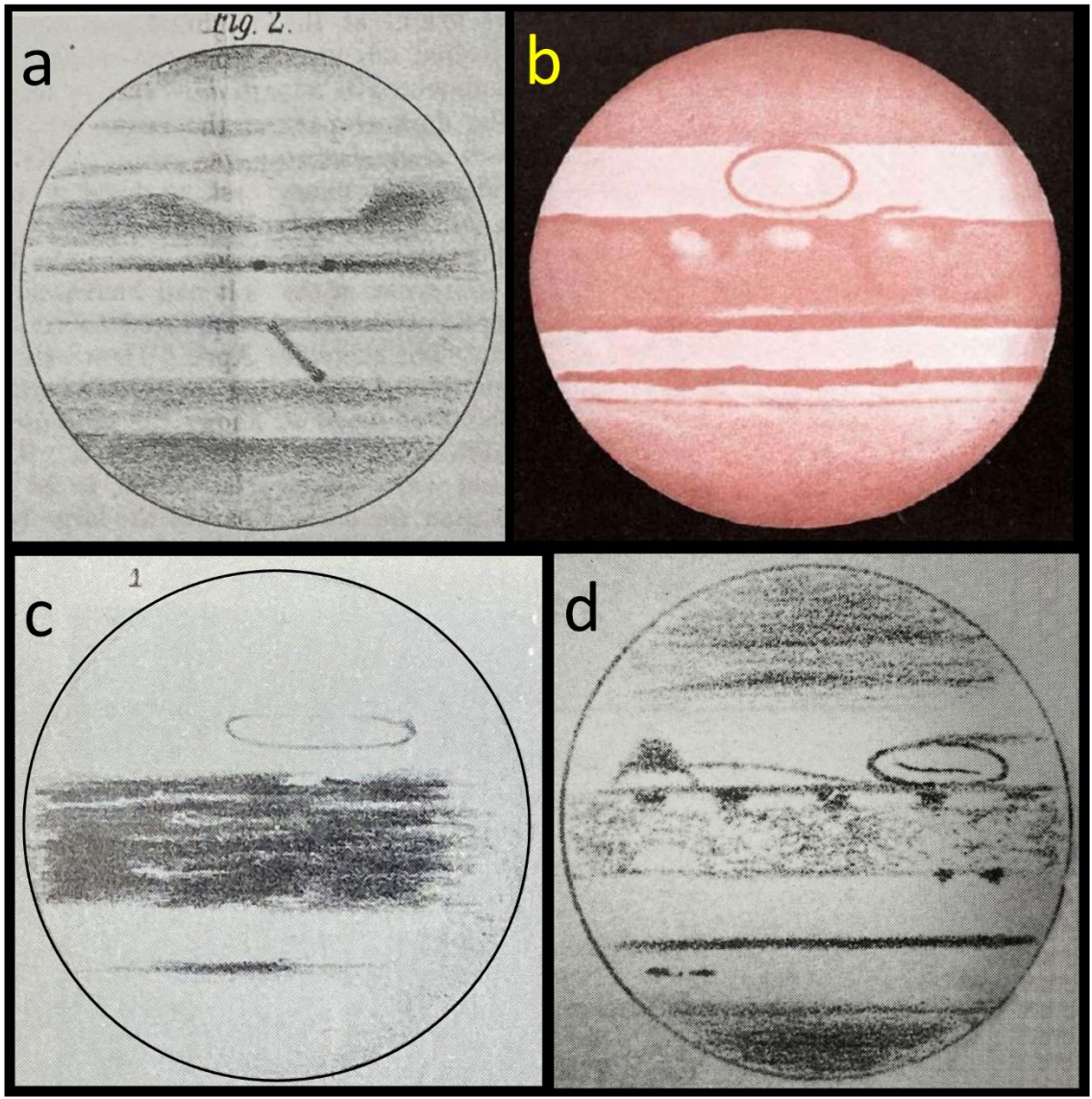

**Figure S3.** The GRS in 1860-1871. (a) Hollow aspect of the GRS in a drawing by J. Baxendell, 2 March 1860; (b)-(d) Drawings showing the GRS as a clear ellipse traced by dark ring. (b) A. M. Mayer in 5 January 1870; (c) M. Mitchell in 1870; (d) J. Gledhill in 1 December 1871.

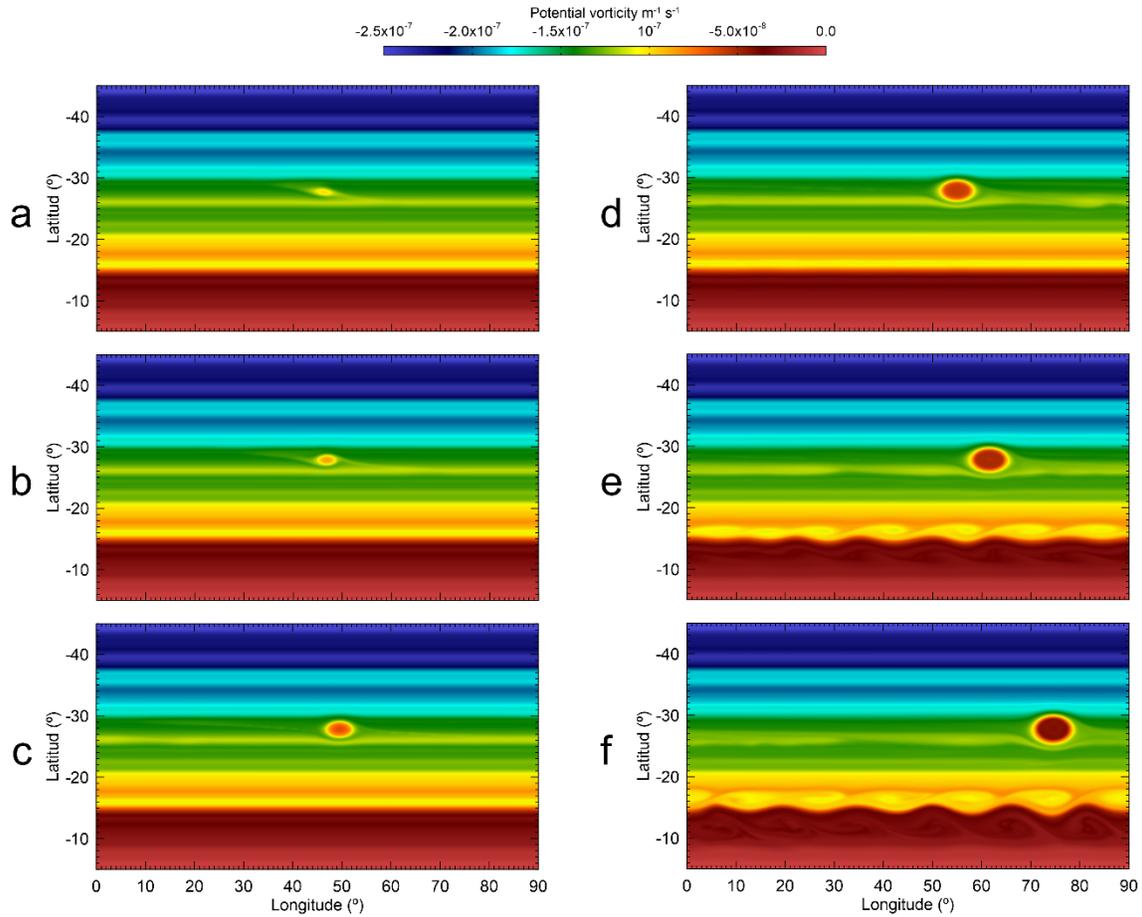

**Figure S4.** Generating an anticyclone from a super-storm simulation in a Shallow Water (SW) model. Maps of the evolution of the potential vorticity field following a mass injection of $10^{10}$ m$^3$s$^{-1}$ introduced during 10 days. The injection takes place in a Gaussian area with 1° radius (the $\sigma$ value of the Gaussian) and limited to a range of 3° at planetographic latitude 23°S. Simulation times (days): (a) 5, (b) 8, (c) 20, (d) 40, (e) 60, (f) 90. The generated anticyclone has a length to width 9° times 5°.

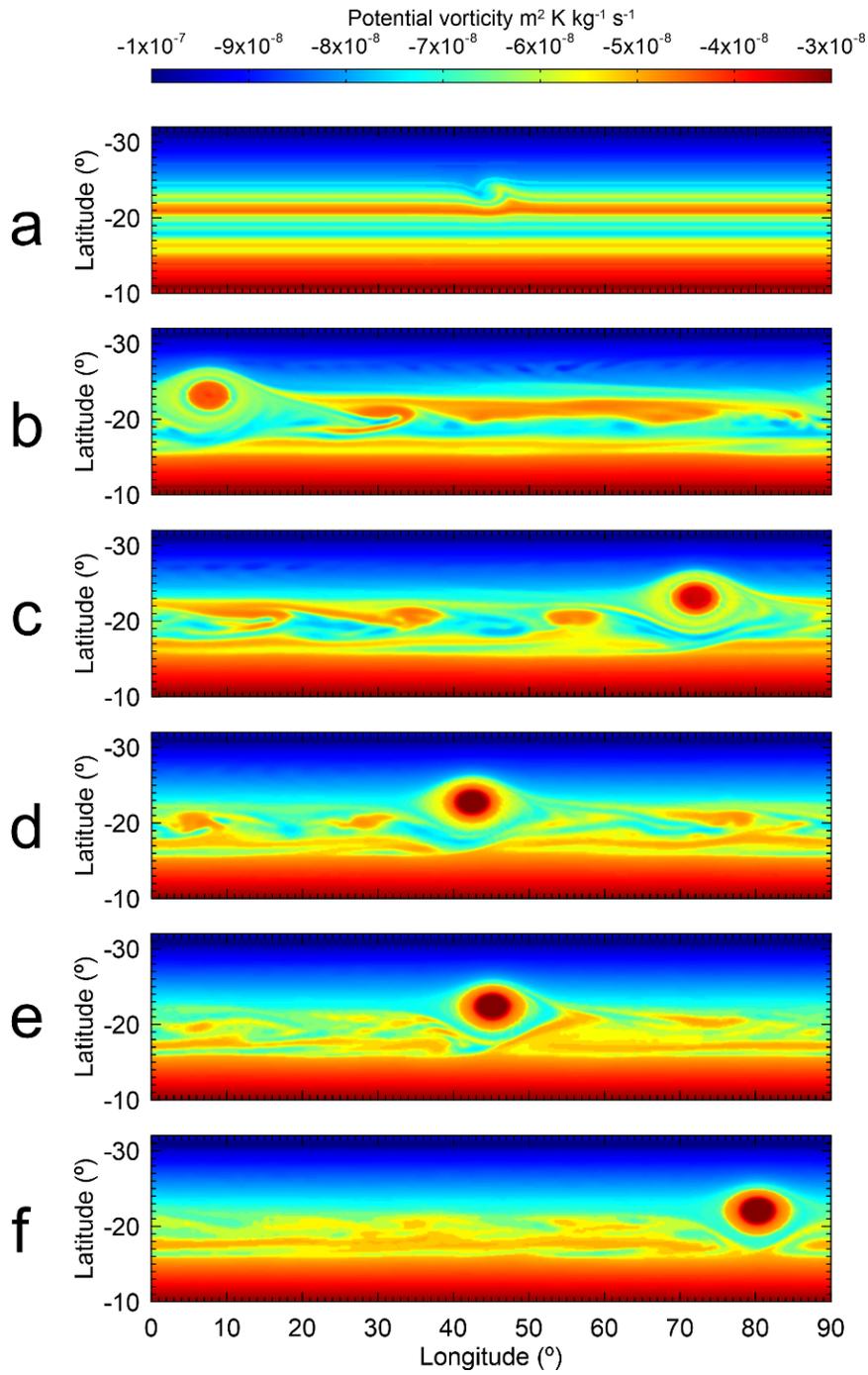

**Figure S5.** Generating an anticyclone from a Super-storm simulation in the EPIC model. Maps of the evolution of the potential vorticity field following a heat impulse with a Gaussian shape with size 0.5° and intensity of 1.5 Wkg$^{-1}$ introduced during 10 days at latitude 23.7°S. An anticyclone with a length to width 9° times 6° and a Hollow with a size ~ 15° times 9° forms after 400 days of simulation. The time sequence (days) is: (a) 1, (b) 50, (c) 100, (d) 200, (e) 300, (f) 400.

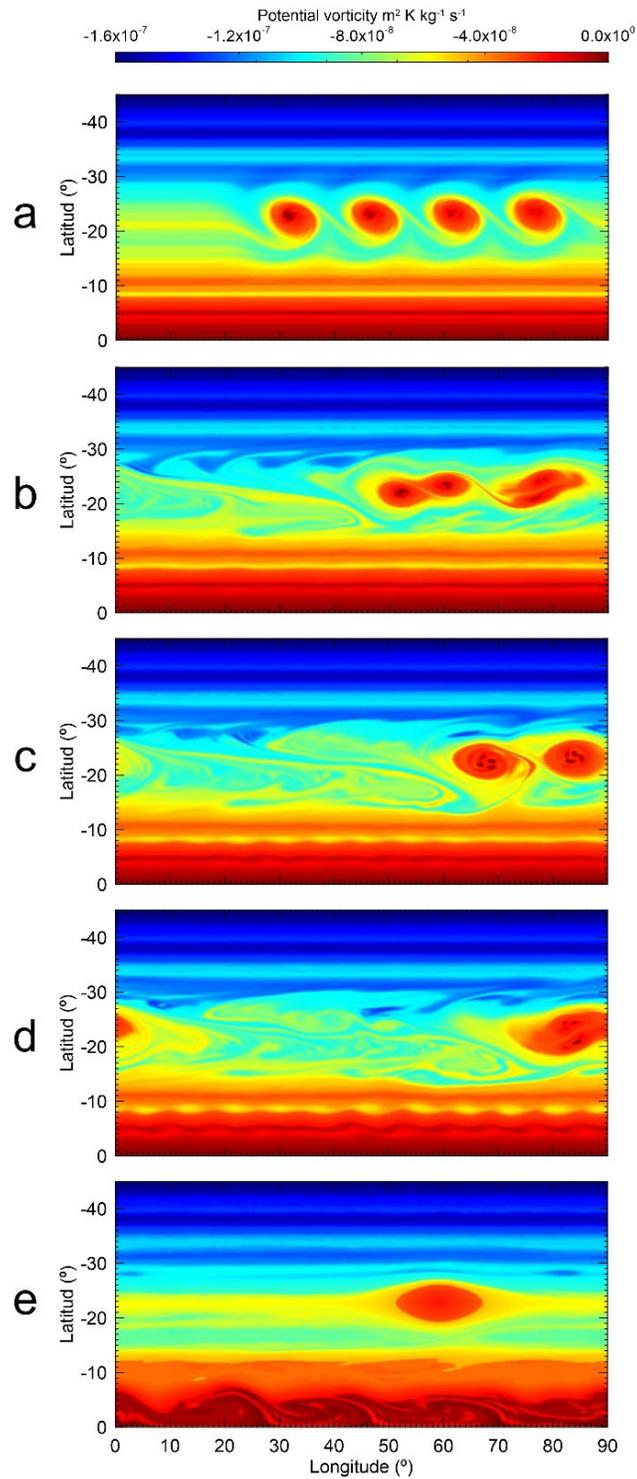

**Figure S6.** Generating a large anticyclone from mergers of vortices in the EPIC model. PV maps of the simulation of the mergers of four anticyclones with an initial size 8° times 7° and periphery velocity 120 ms$^{-1}$ located in slightly different latitudes between 22°S to 22.5°S to force their mutual interaction. The time sequence (days) is: (a) 1, (b) 9, (c) 15, (d) 20, (e)125.

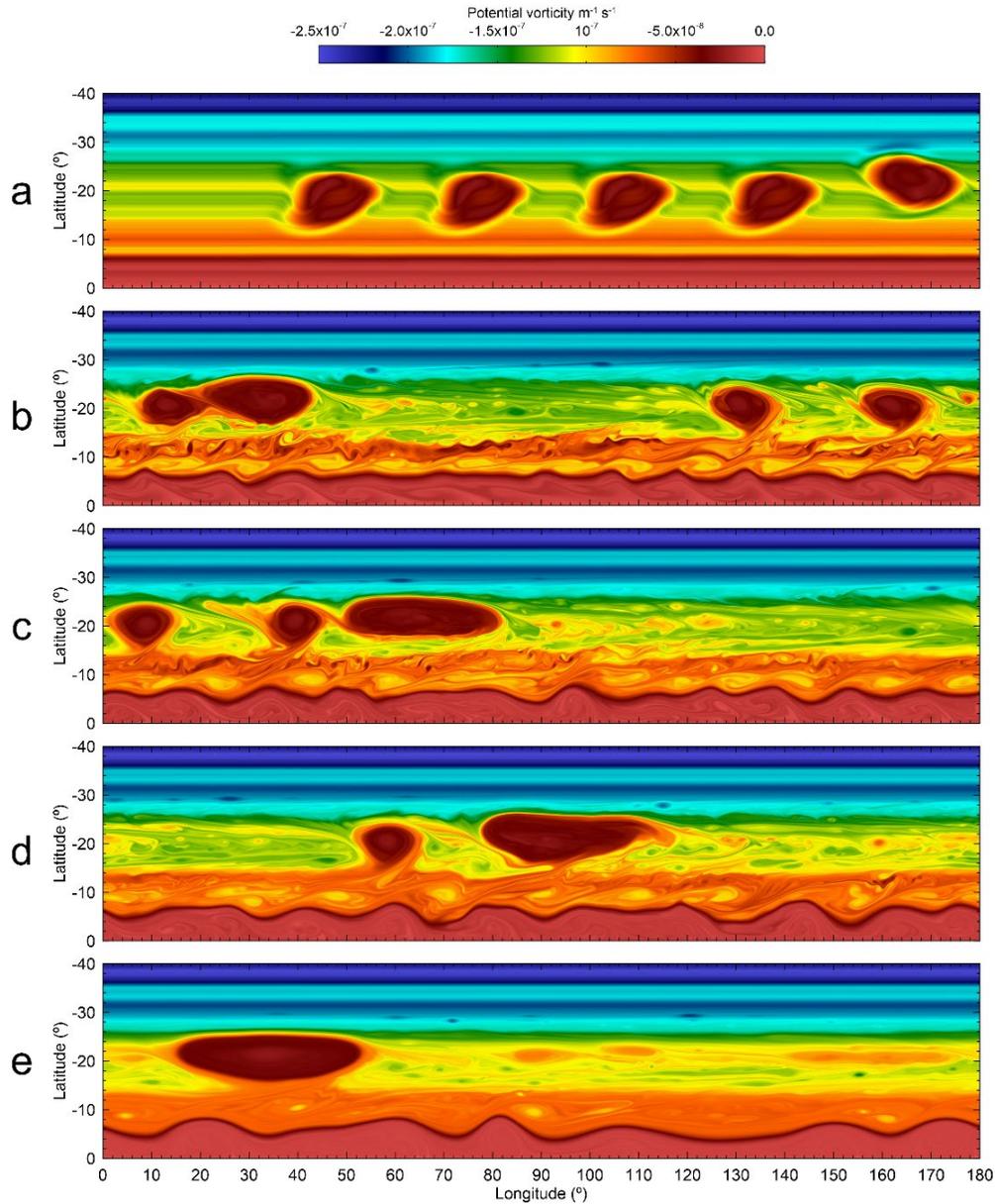

**Figure S7.** Generating a large anticyclone from mergers of vortices in the SW model. Sequence PV fields for mergers of five anticyclones with a maximum tangential velocity of 90 ms$^{-1}$ as a Gaussian perturbation with a major axis of 15° and a minor axis of 10°. One of the vortices was introduced at a latitude of 22.5°S whereas the other four where injected at 24°S. The latitude difference was used to forced a different drift velocity and quickly produce the merging process. Panels correspond to the following simulation days: (a) 1, (b) 38, (c) 62.5, (d) 84, and (e) 250. The resulting single anticyclone has a size 41° times 12°.

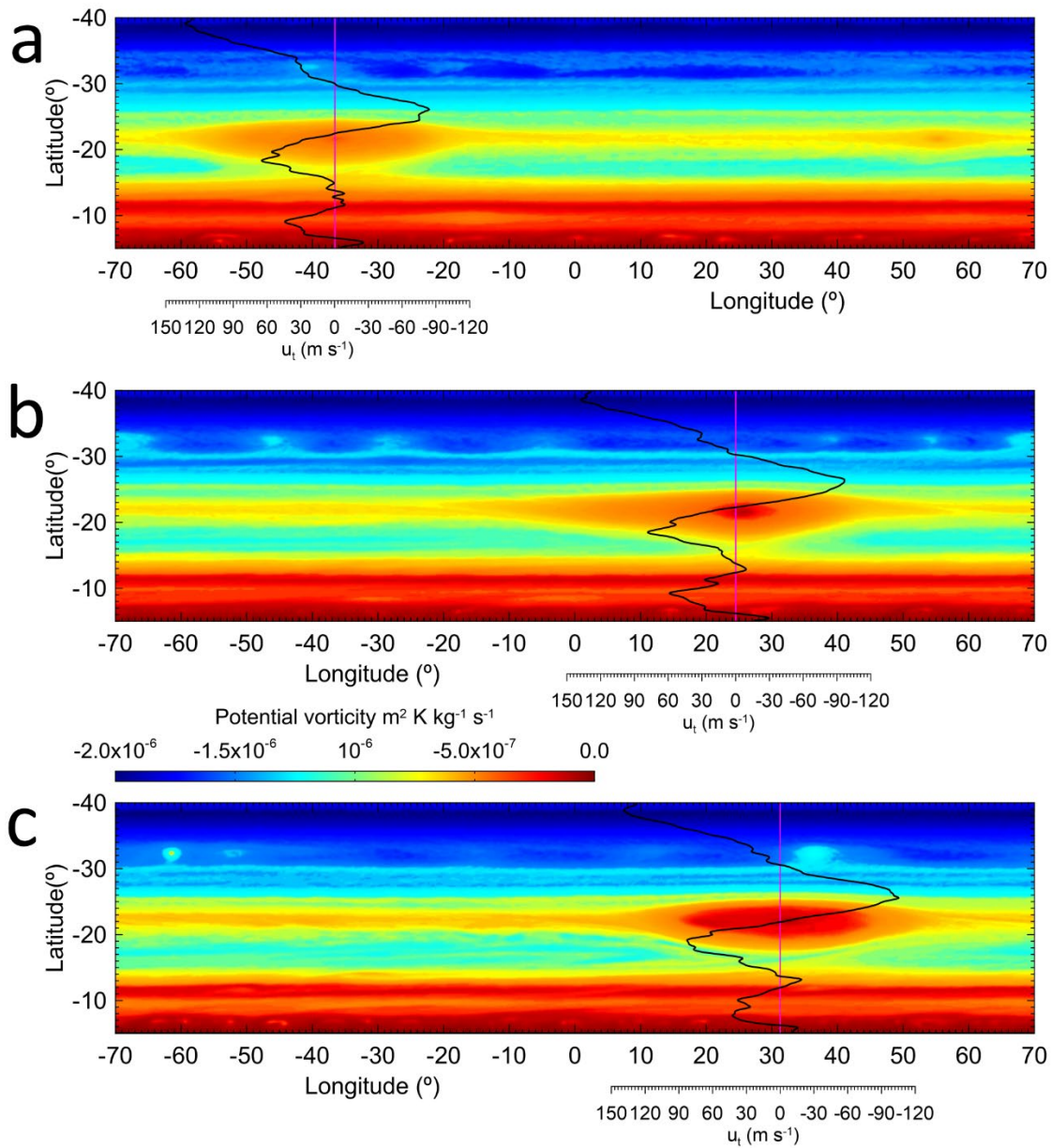

**Figure S8.** Simulations of the evolution of elongated cells in EPIC model. PV maps in the pressure layer 700 mbar (the location of the upper visible clouds) showing the evolution of narrow and long anticyclone cells centred at latitude 22.5°S. The initial length to width of the cell is 30° times 6.5° for different periphery velocities: (a) 50 ms$^{-1}$, time simulation 400 days; (b) 75 ms$^{-1}$, time simulation 385 days; (c) 100 ms$^{-1}$, time simulation 360 days. Superimposed on the oval, the black line represents the zonal wind speed in a North-South cut through the centre of the oval.

**Table S1**. Sources for the images and data plotted in Figure 2.

| Period (years) | Studied Features | Sources | Document |
|---|---|---|---|
| 1631-1714 | PS (Permanent Spot) | 1 | Drawings |
| 1714-1831 | No spot reports | 2 | Drawings |
| 1831-1880 | Early GRS | 3 | Drawings |
| 1890-1915 | GRS | 4 | Photographs |
| 1915-1971 | GRS | 4-5 | Photographs |
| 1963-1972 | GRS | 6 | Photographs |
| 1973-2023 | GRS | 7 | Photographs & digital |
| 1991-2023 | GRS | 8 | HST digital imaging |

**1. Years 1631-1714**

References in the paper: Rogers (1995), Cassini (1666), Chapman (1968), Falorni (1987), Hockey (1999), Simon (2016), Chapman (2016), Graney (2010), Cassini (1692), Maraldi (1708), Cassini (1672), Johns (1992)

Donato Creti, Astronomical Observations, Musei Vaticani
https://m.museivaticani.va/content/museivaticani-mobile/en/collezioni/musei/la-pinacoteca/sala-xv---secolo-xviii/donato-creti--osservazioni-astronomiche.html

Manfredi E. (on the paints by Donato Creti). Linda Hall Library.
https://www.lindahall.org/about/news/scientist-of-the-day/eustachio-manfredi

**2. Years 1714-1831**

References in the paper: Rogers (1995), Hockey (1999), Messier (1769), Herschel (1781), Dobbins et al. (1997)

**3. Years 1831-1880**

References in the paper: Peek (1958), Rogers (1995), Hockey (1999), Clerke (1887)

Secchi A. Memorie dell Osservatorio del Collegio Romano 1852-55, plates, Roma (1856)

Dawes, W. Further observations of the round spots on one of the Belts of Jupiter, *Monthly Notices of the Royal Astronomical Society,* **18**, 49-50 and plates (1857)

Airy G. Remarks on the appearance of Jupiter, *Monthly Notices of the Royal Astronomical Society,* **20**, 244-245 (1860). Drawings by J. Long, J. Baxendell and Mr. Fletchers in plates.

**4. Years 1890-1975**

A large part of the images comes from the digital archive of planetary images at:
Base de Données d'Images Planétaires (BDiP) at Paris-Meudon Observatory.
BDIP - Observatoire de Paris (accessed 2024):
http://www.lesia.obspm.fr/BDIP/bdip.php)

**5. Years 1915-1971**

References in the paper: Peek (1958), Rogers (1995)

Lick Observatory Records, UC Santa Cruz
https://digitalcollections.library.ucsc.edu/collections/tx31qn116

Beebe R. F., Orton G. S., & West R. A. Time Variable Nature of the Jovian Cloud Properties and Thermal Structure: An Observational Perspective, in *Time-Variable Phenomena in the Jovian System,* M. J. S. Belton, R. A. West, J. Rahe (editors), NASA SP-494, 245-288 (1989)

Slipher E. C. Photographs of Jupiter, 1915-1940, *Lowell Observatory Archives* (accessed 2023)
https://collectionslowellobservatory.omeka.net/items/show/1169

See also "A Photographic Study of the Brighter Planets", Lowell Observatory, Flagstaff, Arizona and the National Geographic Society, Washington D. C. (1964)

The University of Chicago Photographic Archive
https://cphotoarchive.lib.uchicago.edu/db.xqy?keywords=Jupiter

Humason M. L. Photographs of the Planets with the 200-inch Telescope, in Planets and Satellites, edited by G. P. Kuiper and B. M. Middlehurst, Vol III, p. 572 (plates) (1961).

Communications of Lunar and Planetary Laboratory Vol. 9, Part 5, Communications Nos. 173-183, The University of Arizona (1972-73)

**6. Years 1963-1972**

References in the paper: Peek (1958), Rogers (1995).

Reese E. J. & Solberg H. S. Recent measures of the latitude and longitude of Jupiter's Red Spot, *Icarus* **5**, 266-273 (1966)

Solberg H. S. Jupiter's Red Spot in 1965-1966, *Icarus* 8, 82-89 (1968a)

Solberg H. S. Jupiter's Red Spot in 1966-1967, *Icarus* 9, 212-216 (1968b)

Solberg H. S. Jupiter's Red Spot in 1967-1968, *Icarus* 10, 412-416 (1969)

Reese E. J. Jupiter's Red Spot in 1968-1969, *Icarus* 12, 249-257 (1970)

Reese E. J. Jupiter: Its Red Spot and other features in 1969-1970, *Icarus* 14, 343-354 (1971)

Reese E. J. Jupiter: Its Red Spot and disturbances in 1970-1971, *Icarus* 17, 57-72 (1972)

Inge J. L. Short-Term Jovian Rotation Profiles. 1970-1972, *Icarus*, 29, 1-6 (1973).

**7. Years 1973-2023**

References in the paper: Rogers (1995), Simon et al. (2018)

Isao Miyazaki: Jupiter images from Okinawa Island (1988-2023)
http://www.ii-okinawa.ne.jp/people/miyazaki/planet/

Association of Lunar and Planetary Observers (ALPO Japan) (1999-2023)
http://alpo-j.sakura.ne.jp/indexE.htm

Planetary Virtual Observatory Laboratory (2000-2023)
http://pvol2.ehu.eus/pvol2/

Association of Lunar and Planetary Observers (USA), Gallery period 1962-1966
https://alpo-astronomy.org/gallery3/index.php/Jupiter-Images-and-Observations

Société Astronomique de France, Commision des Observations Planétaires, section Jupiter (1992-2023)
http://www.astrosurf.com/planetessaf/jupiter/

British Astronomical Association (BAA)
http://britastro.org

**8. Years 1991-2023**

HST (Hubble Space Telescope)
NASA Planetary Data System (PDS) (OPUS3)
https://opus.pds-rings.seti.org/opus/#/mission=Hubble&cols=opusid,instrument,planet,target,time1,observationduration&widgets=mission&order=time1,opusid&view=search&browse=gallery&cart_browse=gallery&startobs=1&cart_startobs=1&detail=

Outer Planet Atmospheres Legacy (OPAL)
https://archive.stsci.edu/hlsp/opal

**Table S2.** Parameters used in EPIC numerical simulations.

The vertical profiles of temperature, Brun-Väisälä frequency and vertical shear of the zonal wind profile, are the same than those given in Figure S4 of Supporting Information in Sánchez-Lavega et al. (2021).

General data

| | |
|---|---|
| Domain extent (deg) | Longitude = 0°-135°, 0°-270°; Latitude = 5°S-45°S |
| Resolution (deg pix$^{-1}$) | 0.17 |
| Time step (s) | 30 |
| Number of layers | 10 |
| Layer limits | 10 mbar (top) – 20 bar (bottom) |
| Hyperviscosity ($\nu_6$, m$^6$s$^{-1}$) | 0.8 x10$^{26}$ |

Superstorm simulations

| | |
|---|---|
| Central latitude | 23.5°S |
| Gaussian initial size | 0.5° - 1° |
| Gaussian heat injection | 0.5 - 10 W kg$^{-1}$ |
| Injection time | 1, 5, 10 days |
| Injection layers | 3 – 7 (1.4 – 7 bar) |
| Simulation time | 500 days |

Vortices mergers simulations

| | |
|---|---|
| Number of anticyclones | 2 - 4 |
| Latitude location Planetographic (south) | 22.5°, 22.7°, 22.9°, 23.1° |
| Gaussian vortex size radius | 5° (longitude) times 4° (latitude) |
| Peripheral velocity | 120 ms$^{-1}$ |
| Reference Pressure level (P0) (mbar) | 1000 |
| Altitude range (top, scale heights) | 3 |
| Altitude range (down, scale heights) | 2.3 |
| Simulation time | 500 days |

STrD mechanism: long-cells

| | |
|---|---|
| Initial circulating cell size | Longitude = 60°, Latitude = 13° |
| Latitude location Planetographic (south) | 22.5° |
| Peripheral velocity | 50 – 100 ms$^{-1}$ |
| Simulation time | 300 days |

**Table S3.** Parameters used in the Shallow Water (SW) numerical simulations.

General data

| Domain extent (deg) | Longitude = 0° - 180°, Latitude = 5° - 45° |
|---|---|
| Resolution (deg/pixel) | 0.02 – 0.2 |
| Time step (s) | 0.25 - 10 |
| Layer thickness (m) | 1000 |
| Simulation time (days) | 150 – 450 |

Superstorm simulations

| Central latitude | 22.5°S |
|---|---|
| Gaussian initial size | 2° - 14° |
| Gaussian heat injection | $10^9 - 10^{12}$ m$^3$ s$^{-1}$ |
| Injection time | 10 - 100 days |
| Simulation time | 300 days |

Vortices mergers simulations

| Number of anticyclones | 5 |
|---|---|
| Latitude location | One at -22.5° – Rest at -19º, -22°, -24º |
| Gaussian vortex size | 15° (longitude) times 10° (latitude) |
| Peripheral velocity | 120 ms$^{-1}$ |
| Simulation time | 250 days |

STrD mechanism: long-cells

| Initial circulating cell size | Longitude = 45° - 80°, Latitude = 13° |
|---|---|
| Latitude location | 22.5° |
| Peripheral velocity | 50 – 100 ms$^{-1}$ |
| Simulation time | 425 days |